\documentclass[aps,twocolumn,groupedaddress]{revtex4}
\usepackage{amssymb}
\usepackage{graphicx}
\usepackage{dcolumn}
\usepackage{bm}

\begin{document}

\title{Electron transport in a slot-gate Si MOSFET}

\author{I. Shlimak}
\affiliation{Jack and Pearl Resnick Institute of Advanced Technology,
Department of Physics, Bar-Ilan University, Ramat-Gan 52900, Israel}

\author{V. Ginodman}
\affiliation{Jack and Pearl Resnick Institute of Advanced Technology,
Department of Physics, Bar-Ilan University, Ramat-Gan 52900,  Israel}

\author{A. Butenko}
\affiliation{Jack and Pearl Resnick Institute of Advanced Technology,
Department of Physics, Bar-Ilan University, Ramat-Gan 52900,  Israel}

\author{K.-J. Friedland}
\affiliation{Paul-Drude Institut f\"ur Festk\"orperelektronik,
Hausvogteiplatz 5-7, 10117, Berlin, Germany}

\author{S. V. Kravchenko}
\affiliation{Physics Department, Northeastern University, Boston, Massachusetts 02115, U.S.A}

\begin{abstract}
The transversal and longitudinal resistance in the quantum Hall effect regime 
was measured in a Si MOSFET sample in which a slot-gate allows one to vary 
the electron density and filling factor in different parts of the sample.  
In case of unequal gate voltages, the longitudinal resistances on the opposite sides 
of the sample differ from each other because the originated Hall voltage difference 
is added to the longitudinal voltage only on one side depending on the gradient 
of the gate voltages and the direction of the external magnetic field. 
After subtracting the Hall voltage difference, the increase in longitudinal 
resistance is observed when electrons on the opposite sides of the slot occupy 
Landau levels with different spin orientations.
\end{abstract}


\maketitle

\section{Introduction}

The fabrication of Si-MOSFET samples with a narrow gate barrier or with
narrow slots in the gate~\cite{Haug,Washburn,Berkut,Wang,vanSon} has given
rise to new experimental possibilities. In particular, samples with a narrow
gate barrier~\cite{Haug,Washburn} were used for investigation of the
backscattering of the edge current in the quantum Hall effect (QHE) regime,
while the slot-gate geometry has permitted reliable measurements of a
two-dimensional electron transport in case of low electron density~\cite{Shashkin}. 
In this work, we use the slot-gate geometry to measure the
longitudinal resistance $R_{xx}$ in the QHE regime for unequal electron
densities along the sample. Our aim was to reveal the influence of the
spin-flip process on the electron transport when electrons on the opposite
sides of the slot occupy Landau levels (LL) with different spin orientations.

\section{Experimental results and discussion}

\begin{figure}[t]
\includegraphics{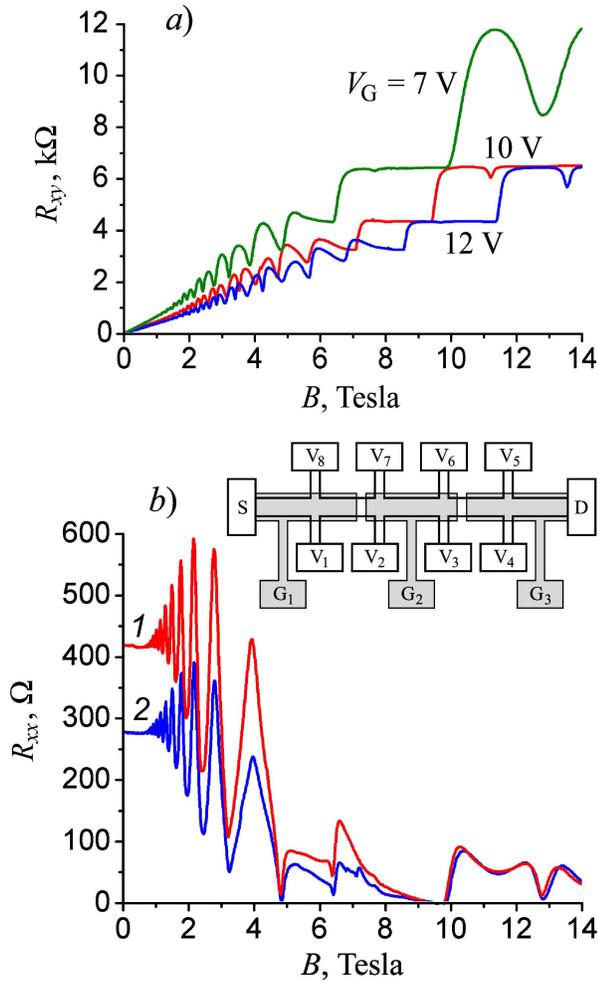}
\caption{\emph{a} --- transverse resistance $R_{xy}$ as a function of
magnetic field at fixed gate voltages $V_\mathrm{G}=7, 10, 12$\,V; \emph{b}
--- longitudinal resistance $R_{xx}$ measured between probes $\mathrm{V}_1$
and $\mathrm{V}_2$, $R_{12}$ (\textsf{\textit{1}}) and between probes
$\mathrm{V}_2$ and $\mathrm{V}_3$, $R_{23}$ (\textsf{\textit{2}}) at
$V_\mathrm{G}=7$\,V. The insert shows the schematics of the slot-gate sample.}
\label{Fig1}
\end{figure}

The sample with two narrow slots (100~nm) in the upper
metallic gate was similar to that described earlier in Ref.~\cite{Shashkin}
(see insert in fig.~\ref{Fig1}).
Application of different gate voltages $V_\mathrm{G}$ to the gates 
$\mathrm{G}_1$, $\mathrm{G}_2$ and $\mathrm{G}_3$ permitted one to maintain
different electron densities $n$ in different parts of the sample.

The sample resistance was measured at $T=40$\,mK using a standard lock-in
technique with the measuring current 20~nA at a frequency of
10.6~Hz. The electron mobility was $\mu =2.68\mathrm{\,m^2/V\cdot s}$
at $n=0.83\cdot 10^{16}$\,m$^{-2}$.

In the first series of experiments, all gates were connected. The magnetic 
field dependences of the Hall (transverse) resistance $R_{xy}$ measured 
between probes $\mathrm{V}_{2}$--$\mathrm{V}_{7}$ are shown in 
fig.~\ref{Fig1}\emph{a}. The ``plateaus'' are clearly seen only for Landau filling
factors $\nu \equiv hn/eB=4$ and $\nu =6$ corresponding to the Hall
resistances $6.45\mathrm{\,k\Omega}=1/4(h/e^{2})$ 
and $4.3\mathrm{\,k\Omega}=1/6(h/e^{2})$, correspondingly. 
Clear ``plateaus'' in $R_{xy}$ at $\nu >6$ are usually
not observed in Si-MOSFET being ``contaminated'' by the ``overshoot'' effect 
\cite{Richter,Shlimak1}.

The longitudinal resistance $R_{xx}$ was measured across the gap between
voltage probes $\mathrm{V}_{1}$ and $\mathrm{V}_{2}$ $(R_{12})$ and without
the gap, between probes $\mathrm{V}_{2}$ and $\mathrm{V}_{3}$ $(R_{23})$. In
zero magnetic field, $R_{12}$ is 1.5 times larger than $R_{23}$ 
(fig.~\ref{Fig1}\emph{b}) due to the distance between probes 1 and 2 being 1.5 times
larger than that between probes 2 and 3. Therefore, the longitudinal
resistance is not affected by the existence of the narrow slot in the gate.
In other words, in our sample, the narrow slot in the upper gate does not
lead to the existence of a potential barrier.
It is remarkable, however, that at magnetic fields above 10\,T, both
resistance curves merge. This can be explained by the influence of the edge channels 
\cite{edge,edge1}, so that the length between probes becomes irrelevant.

\begin{figure}[t]
\includegraphics{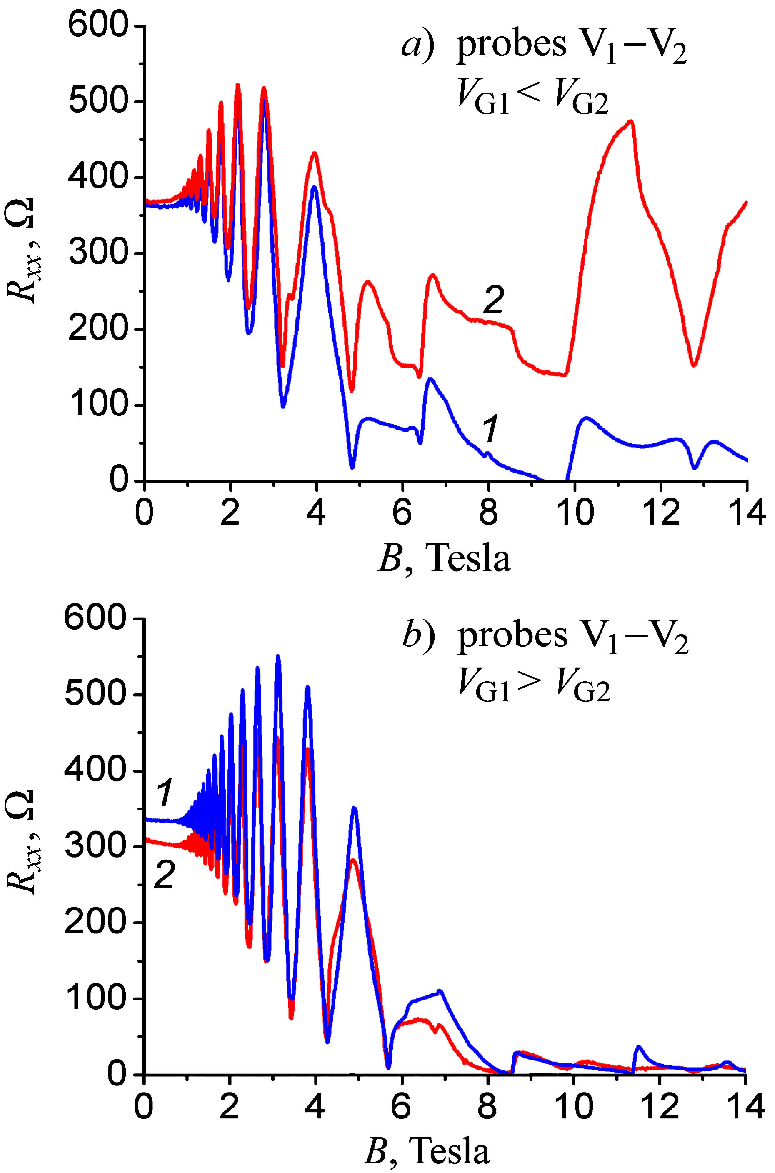}
\caption{Longitudinal resistance $R_{12}$ measured when $V_{\mathrm{G}1}\neq
V_{\mathrm{G}2}$. \emph{a}~--- $R_{12}$ measured when 
$V_{\mathrm{G}1}<V_{\mathrm{G}2}$ ($V_{\mathrm{G}1}=7\mathrm{\,V}$, 
$V_{\mathrm{G}2}=12\mathrm{\,V}$) (curve~\textsf{\textit{2}}); $R_{12}$ for the case 
of equal gate voltages $V_{\mathrm{G}1}=V_{\mathrm{G}2}=7\mathrm{\,V}$ is shown 
for comparison (curve~\textsf{\textit{1}}); 
\emph{b}~--- $R_{12}$ measured when 
$V_{\mathrm{G}1}>V_{\mathrm{G}2}$ ($V_{\mathrm{G}1}=12\mathrm{\,V}$, 
$V_{\mathrm{G}2}=7\mathrm{\,V}$) (curve~\textsf{\textit{2}}), $R_{12}$ for the case 
of equal gate voltages $V_{\mathrm{G}1}=V_{\mathrm{G}2}=12\mathrm{\,V}$ is shown 
for comparison (curve~\textsf{\textit{1}}).}
\label{Fig2}
\end{figure}

In case of different gate voltages $V_{\mathrm{G}1}\neq V_{\mathrm{G}2}$ 
($V_{\mathrm{G}3}$ was always equal to $V_{\mathrm{G}2}$), the difference in
the transverse Hall voltages $\Delta V_{\mathrm{H}}\equiv V_{18}-V_{27}$
appears. $\Delta V_{\mathrm{H}}$ is added to the longitudinal voltage $V_{xx}
$ only on one side of the sample, depending on the gradient of the gate
voltage $\nabla V_{\mathrm{G}}$, which makes the longitudinal resistance
non-symmetric: for $V_{\mathrm{G}1}<V_{\mathrm{G}2}$ at given direction of
the magnetic field $\mathrm{\textbf B}$, $\Delta V_{\mathrm{H}}$ was added to the
voltage $V_{12}$, while for $V_{\mathrm{G}1}>V_{\mathrm{G}2}$, $V_{12}$
remains unchanged (fig.~\ref{Fig2}\emph{a,b}). On the opposite side of the
sample, the situation is reverse: $\Delta V_{\mathrm{H}}$ is added to the
voltage drop $V_{87}$ for $V_{\mathrm{G}1}>V_{\mathrm{G}2}$. It was shown in
Ref.~\cite{Berkut} that the sample side where $\Delta V_{\mathrm{H}}$ is
added to $V_{xx}$ is determined by the vector product 
$\mathrm{\textbf B}\times\nabla V_{\mathrm{G}}$. Different values 
of $V_{xx}$ on the opposite sides of the
sample mean that in order to analyse the longitudinal resistance in the case
of different gate voltages, one need to subtract properly the contribution
of the Hall voltage difference.

\begin{figure*}[t]
\parbox[b]{8.5cm}{
\includegraphics[width=8cm]{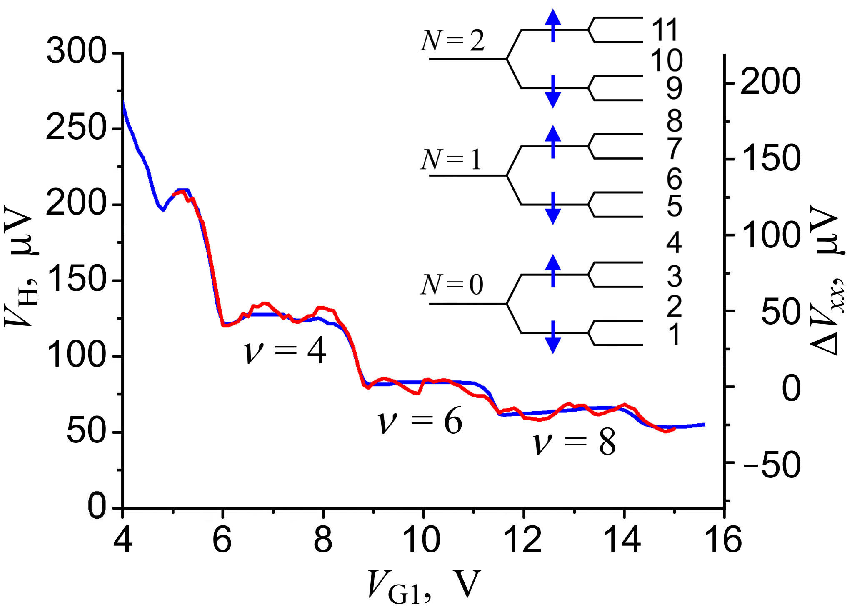}
\caption{Dependence of the Hall voltage $V_\mathrm{H}$ on the gate voltage
at fixed magnetic field $B=8$\,T (blue line, left scale). The inset
shows distribution of Landau levels in a Si-MOSFET with cyclotron, spin and
valley splittings indicated. Numbers correspond to the integer values of the
filling factor $\nu$. The difference of the longitudinal voltages
measured on the opposite sides of the sample $\Delta V_{xx}=V_{87}-V_{12}$
is plotted as a function of $V_{\mathrm{G}1}$ at fixed 
$V_{\mathrm{G}2}=10$\,V for comparison (red line, right scale).}\vspace*{6mm}
\label{Fig3}}
\hfill
\parbox[b]{8.5cm}{
\includegraphics[width=6.8cm]{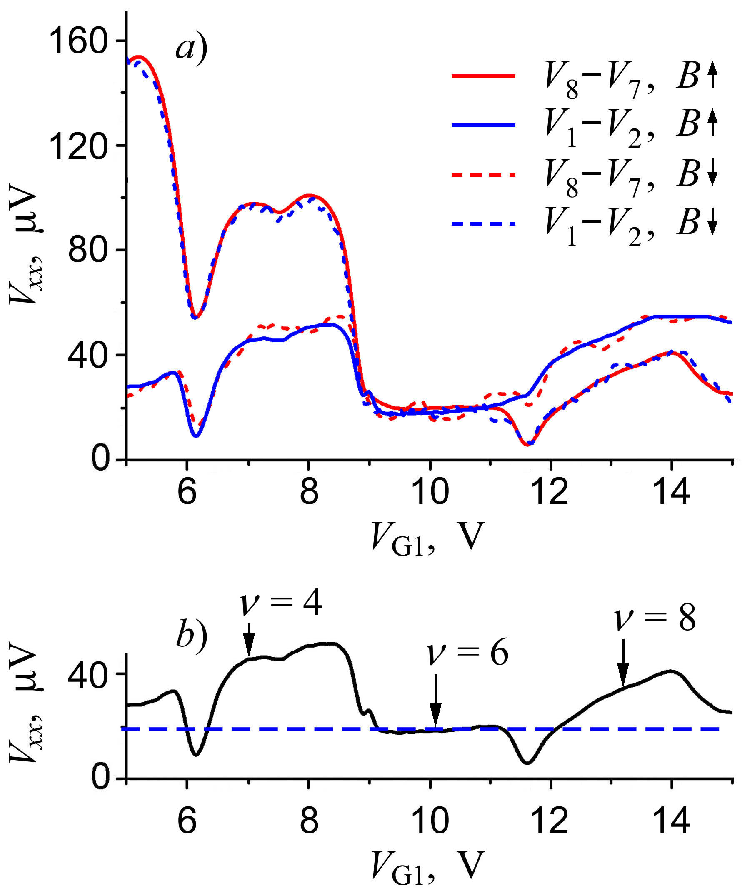}
\caption{\emph{a} --- longitudinal voltages across the slot $V_{87}$ and
$V_{12}$ as a function of $V_{\mathrm{G}1}$ at fixed
$V_{\mathrm{G}2}=10$\,V and $B=8$\,T,
\emph{b} --- lower part of the curves: no contribution of
the Hall voltage difference.}
\label{Fig4}}
\end{figure*}

In another set of experiments, conducted at $T=300$\,mK, the magnetic
field was fixed at 8~Tesla, while the gate voltage was varied. First, all
gates were connected and $V_{\mathrm{H}}$ was measured between probes 
$\mathrm{V}_{1}$ and $\mathrm{V}_{8}$ (figure~\ref{Fig3}, blue line). Using
the data shown in fig.~\ref{Fig1}\emph{a}, one can conclude that the
``plateau'' at around $V_{\mathrm{G}}=10$\,V corresponds to the 
filling factor $\nu =6$, while the ``plateaus'' at around 
$V_{\mathrm{G}}=7$\,V and $V_{\mathrm{G}}=13$\,V correspond to 
$\nu =4$ and $\nu =8$, correspondingly.

Figure~\ref{Fig4}\emph{a} shows $V_{xx}$ measured simultaneously on both
sides of the sample between probes $\mathrm{V}_{1}$--$\mathrm{V}_{2}$ and
probes $\mathrm{V}_{8}$--$\mathrm{V}_{7}$.
$V_{\mathrm{G}2}=10\mathrm{\,V}$ was kept constant, while $V_{\mathrm{G}1}$ was
varied from 5\,V to 15\,V. In this experiment, the gradient of the
gate voltage undergoes a sign change at $V_{\mathrm{G}1}=10\mathrm{\,V}$. If the
direction of the magnetic field is reversed, the curves trade places. The
difference between the two curves $\Delta V_{xx}$ plotted as a function of 
$V_{\mathrm{G}1}$ (fig.~\ref{Fig3}, red line) practically coincides with 
$V_{\mathrm{H}}(V_{\mathrm{G}1})$. This fact allows us to subtract properly
the contribution of the Hall voltage difference: one need to take into
account only the lower parts of the both curves. The result is shown in 
fig.~\ref{Fig4}\emph{b}.

Let us discuss the curve shown in fig.~\ref{Fig4}\emph{b}. Keeping constant 
$V_{\mathrm{G}2}=10\mathrm{\,V}$ means that electrons underneath the gate $\mathrm{G}_2$ 
always occupy the sixth LL with spin ``down'' (see fig.~\ref{Fig3} and
the inset). When $V_{\mathrm{G}1}$ is varied from 9\,V to 11\,V,
electrons across the slot occupy the same (6th) LL and, therefore, have the
same spin orientation. However, when 7\,V$<V_{\mathrm{G}1}<$9\,V and 
12\,V$<V_{\mathrm{G}1}<$14\,V, the electrons underneath the gate 
$\mathrm{G}_1$ occupy ``spin up'' LLs 4 and 8, correspondingly. At both
filling factors $\nu=4$ and $8$ (indicated in the figure by arrows), 
the longitudinal resistance increases compared to the 
$\nu=6$ case. Possible reason for this resistance increase is some 
additional scattering~\cite{Buttiker,Vakili} due to the necessity for 
electrons to flip their spins when crossing the slot.

\acknowledgments I. S. thanks V. T. Dolgopolov and A. A. Shashkin for
fruitful discussion and the Erick and Sheila Samson Chair of Semiconductor
Technology for financial support. We are grateful to A. Bogush
and A. Belostotsky for assistance.


\begin{thebibliography}{99}
\bibitem{Haug}
R. J. Haug, A. H. MacDonald, P. Streda, and K. von Klitzing,
Phys. Rev. Lett. \textbf{61}, 2797 (1988).

\bibitem{Washburn}
S. Washburn, A. B. Fowler, H. Schmid, and D. Kern,
Phys. Rev. Lett. \textbf{61}, 2801 (1988).

\bibitem{Berkut}
A. B. Berkut, Yu. V. Dubrovskii, M. S. Nunuparov, M. I. Reznikov,
and V. I. Taly’anskii,
JETP Lett. \textbf{44}, 324 (1986).

\bibitem{Wang}
S. L. Wang, P. C. van Son, S. Bakker, and T. M. Klapwijk,
J. Phys.: Condens. Matter \textbf{3}, 4297 (1991).

\bibitem{vanSon}
P. C. van Son, S. L. Wang, and T. M. Klapwijk,
Surface Science \textbf{263}, 284 (1992).

\bibitem{Shashkin}
A. A. Shashkin, S. V. Kravchenko, V. T. Dolgopolov, and T. M. Klapwijk,
Phys. Rev. Lett. \textbf{87}, 086801 (2001).

\bibitem{Richter} 
C. A. Richter, R. G. Wheeler, \and R. N. Sacks,
Surface Science \textbf{263}, 270 (1992).

\bibitem{Shlimak1}
I. Shlimak, V. Ginodman, A. B. Gerber, A. Milner, K.-J. Friedland, and D. J. Paul,
Europhys. Lett. \textbf{69}, 997 (2005).

\bibitem{edge}
P.~L. McEuen, A. Szafer, C.~A. Richter, B.~W. Alphenaar, J.~K. Jain, 
	A.~D. Stone, R.~G. Wheeler, and R.~N. Sacks,
	Phys. Rev. Lett. \textbf{64}, 2062 (1990).
	
\bibitem{edge1}
V.~T. Dolgopolov, G.~V. Kravchenko, and A.~A. Shashkin,
	Solid State Commun. \textbf{78}, 999 (1991).

\bibitem{Buttiker}
M. B\"uttiker,
Phys. Rev. B \textbf{41}, 7906 (1990).

\bibitem{Vakili}
K. Vakili, V. P. Shkolnikov, E. Tutuc, N. C. Bishop, E. P. De Poortere,
and M. Shayegan,
Phys. Rev. Lett. \textbf{94}, 176402 (2005).
\end{thebibliography}
\end{document}